\documentclass[useAMS,usenatbib]{mn2e}
\usepackage{natbib}
\usepackage{graphicx}

\newcommand{\mnras}{MNRAS}
\newcommand{\apj}{ApJ}

\newcommand{\aap}{A\&A}

\newcommand{\physrep}{Phys.\ Reports}
\newcommand{\ssr}{Space Sci.\ Rev.}

\usepackage{natbib}
\usepackage{graphicx}
\usepackage{amsmath}
\usepackage{amssymb}

\newcommand{\eqb}{\begin{equation}}
\newcommand{\eqe}{\end{equation}}
\newcommand{\be}{\begin{eqnarray}}
\newcommand{\ee}{\end{eqnarray}}

\begin{document}
\title{Interaction of the electro-magnetic precursor from a relativistic shock with the upstream flow.
I. Synchrotron absorption of strong electromagnetic waves}

\author[Yuri Lyubarsky]{Yuri Lyubarsky\\
Physics Department, Ben-Gurion University, P.O.B. 653, Beer-Sheva 84105, Israel; e-mail:
lyub@bgu.ac.il}
\date{Received/Accepted}
\maketitle
\begin{abstract}
This paper is the first in  the series of papers aiming to study interaction of the
electro-magnetic precursor waves generated at the front of a relativistic shock with the
upstream flow. It is motivated by a simple consideration showing that the absorption of
such an electro-magnetic precursor could yield an efficient transformation of the kinetic
energy of the upstream flow to the energy of accelerated particles. Taking into account
that the precursor is a strong wave, in which electrons oscillate with relativistic
velocities, the standard plasma-radiation interaction processes should be reconsidered.
In this paper, I calculate the synchrotron absorption of strong electro-magnetic waves.
\end{abstract}

\section{Introduction}

The relativistic wind originating from the rotating, magnetized neutron star (pulsar)
terminates at a strong reverse shock, the shocked plasma inflating within the
surrounding gas a bubble
filled with relativistic particles (mostly electrons and positrons) and magnetic fields. This
bubble is called a pulsar wind nebula (PWNe).
By now, the overall morphology of PWNe is more or less understood in the scope of MHD models (see,
e.g., reviews by \citep{Arons07,Kirk_Lyubarsky_Petri09,Porth_etal17}.
However, 
the physical processes giving rise to particle acceleration in PWNe remain obscure; 
none of the present theories can explain how their spectra formed.

The generic observational feature of PWNe is a flat radio spectrum,
${\cal F}_{\nu}\propto \nu^{-\alpha}$, with $\alpha$ between $0$ and $0.3$, extending in some cases
out to the infrared.  At high frequencies, the spectrum softens, and in the X-ray band, $\alpha>1$.
Such an injection spectrum suggests
a very unusual acceleration process. The observed
radio spectrum
implies a  power-law energy distribution of 
injected electrons, $N(E)\propto E^{-\kappa}$, with a shallow slope $1<\kappa<1.6$. Such an energy
distribution is remarkable in that most of the particles are found at the low energy end of the
distribution, whereas particles at the upper end of the distribution dominate the energy density of
the plasma. Specifically in the Crab Nebula, the observed emission spectrum implies that the
particles in the energy range from $E_{\rm min}<100$ MeV to $E_{\rm break}\sim 1$ TeV are injected
into the nebula with a spectral slope $\kappa=1.6$,
so most of the injected energy ($\sim 5\cdot 10^{38}$ erg$\cdot$s$^{-1}$) is carried by TeV
particles, whereas $\sim 100$ times more particles are found at low energies of less than 100 MeV.
This means that the acceleration process somehow transfers most of the total energy of the system
to a handful of energetic particles, leaving only a small fraction of the energy for the majority
of the particles. This is not what one would normally expect from the conventional first-order
Fermi acceleration process, in which the particle flow is randomized at the shock and only a
fraction of the upstream kinetic energy is deposited in highly accelerated particles.

It was previously assumed \citep{lyub03} that the unusual particle energy distribution in PWNe may
be explained if most of the pulsar spin-down energy is still stored in the striped magnetic field
when the flow enters the termination shock.
In this case, the alternating magnetic fields annihilate at the shock front, and one can speculate
that the radio-to-optical emission of PWNe is generated by pairs accelerated in the course of the
reconnection process.  Particle in cell (PIC) simulations \citep{Sironi_Spitkovsky11a} indeed show
that the alternating fields easily annihilate at the shock. However, nonthermal particle
distributions were found to be generated only if the pair density in the pulsar wind is extremely
high, orders of magnitude larger than that compatible with the observed particle density in the
nebula.
Therefore, an alternative explanation for the unusually flat particle
spectrum in PWNe must be sought.

The aim of this series of papers is to investigate the particle acceleration upstream of the shock
due to absorption of the electromagnetic precursor wave generated at the shock front. The pulsar
wind is magnetized therefore the termination shock is mediated by the Larmor rotation. In this
case, the synchrotron maser instability produces strong, low-frequency electromagnetic waves
propagating both upstream and downstream of the shock \citep{Langdon88,Gallant92,Iwamoto-etal17}
and transferring a few per cent of the upstream energy flow.  A strong precursor wave has also been
found by \citet{amano0kirk13} and \citet{giacche-kirk16} who considered interaction of a circularly
polarized transverse magnetic shear wave, which models the striped structure of the pulsar wind,
with the termination shock. These authors attribute the precursor not to the maser instability but
just to wave conversion at a shock discontinuity. In any case, the energy density of the precursor
wave exceeds, in the comoving frame of the upstream flow, the plasma energy density therefore when
and if the wave is eventually absorbed by the flow, the plasma parameters change significantly even
though in the shock frame, the absorbed energy is small as compared with the flow energy.

  In order to see why this is the
case, consider a body of mass $M$ moving with a high Lorentz factor $\Gamma$ towards a
radiation beam. It follows immediately from energy and momentum conservation that after
the body absorbs some energy $\varepsilon$, it acquires a Lorentz factor
 \eqb
\Gamma_1\simeq\frac{\Gamma}{\sqrt{1+4\Gamma^2\varepsilon/Mc^2}}\, , \eqe
 where $c$ is the
speed of light.  One sees that the body is decelerated significantly if
$\varepsilon>Mc^2/\Gamma^2$. Therefore in the highly relativistic case, the body can
decelerate even if the absorbed energy is small. An observer in the lab frame would say
that most of the kinetic energy of the body has been
transformed into internal energy. 

This simple consideration shows that an electromagnetic precursor can have a profound
effect on the particle acceleration process, because when this radiation is absorbed in
the upstream flow, the kinetic energy of the flow is transformed mostly into internal
energy. The particle spectrum is determined by collisionless absorption processes.
Therefore the internal energy is not thermalized; on the contrary, one would expect
non-thermal particle distributions. As the first step, one has to analyze the decay of a
strong EM wave propagating in a pair plasma.
A few processes look important: synchrotron absorption, induced scattering, three-wave
decay of the pumping wave into an electro-magnetic wave and a magnetosonic wave
(stimulated Brillouin scattering), non-linear self-focusing of the wave. Note that the
Raman scattering of the electro-magnetic wave into another electro-magnetic wave and the
Langmuir wave, does not occur in pair plasmas because the masses of two opposite charges
are equal.
Understanding which of the many processes dominates in  what parameter domain is
essential in order to set up the necessary numerical simulations.

An important point is that the wave is strong in the sense that the wave strength parameter,
 \eqb
a=\frac{eE}{m_ec\omega},
 \eqe
  where $\omega$ and $E$ are the wave  angular frequency and amplitude,
is large. In the field of such a wave, electrons experience oscillations with relativistic
velocities (e.g., \citealp{Landau_Lifshitz75}). Therefore the standard perturbative approach to
plasma-wave interactions could not be used; one has to use methods developed in the field of
laser-matter interaction (e.q., \citealp{mourou06}).  

 There is a vast literature on the interaction of strong waves with plasmas (e.g., reviews by
\citealp{shukla_etal86} and \citealp{mourou06}). However, the parameter range relevant for the case
of interest (the wave frequency is much larger than the plasma frequency so that the wave velocity
is close to $c$, the electron-positron plasma so that effects relied on the mass difference of the
charge carriers, such as Raman scattering, are absent, relativistic mean velocities of particles
etc) has attracted little attention. Radiation of relativistic particles oscillating in a strong
wave (non-linear Compton scattering) has been thoroughly studied
\citep{Gunn_Ostriker71,Arons72,Blandford72,Stewart74}. In particular, a radiative damping of strong
waves has been considered  \citep{Asseo_etal78,Mochol_Kirk13}. However, the spontaneous scattering
could not play a significant role in the case of interest because this process is unable to take a
significant fraction of the flow energy before the flow enters the termination shock (unless a
strong radiation source is presented in the
system, like in PSR B1259-63). 
The induced scattering looks more promising. \citet{Melrose80} derived the kinetic
equation for induced scattering of strong waves, however, the process has not been
 studied thoroughly.

In this paper, I consider synchrotron absorption of strong electromagnetic waves. I
address the high frequency case when the wave propagates like in vacuum; then the
absorption coefficient may be calculated just by finding the average energy the single
electron gains from the vacuum wave. The paper is organized as follows. In the next
section, I present equations of motion for an electron in the presence of a strong
electro-magnetic wave and a background magnetic field. The exact solutions for the zero
background field are reminded and constants of motions of this solution are used as
variables in the case of a weak background field. In sect. 3, the motion of the electron
guiding centre is found in the case when the wave frequency is large as compared with the
Larmor frequency. In sect. 4, small oscillations with respect to the slow Larmor rotation
are considered. In sect. 5, the energy exchange between the wave and the electron is
found and the absorption coefficient is calculated. The validity of the approximations is
analyzed in sect. 6. The obtained results are discussed and qualitatively explained in
sect. 7. In Appendix, the Einstein coefficients method is used to derive the classical
synchrotron absorption coefficient in the weak wave limit.

\section{ Basic equations}

The absorption coefficient may be found by calculating the work done by the wave on the particles.
In the case of a strong wave, one could not consider particle oscillations in the field of the wave
as a small perturbation. However, one could  exploit the fact that the particle motion in the field
of a strong wave may be solved exactly if there is no background magnetic field.
This solution may be used in the presence of the background field if the wave frequency
significantly exceeds the Larmor frequency; then one could find the particle motion
by averaging over the fast wave oscillations.

 The wave generated by the maser instability at the shock front is polarized perpendicularly to
the magnetic field. When the particles in the upstream flow absorb the wave, they begin to rotate
around the magnetic field lines. Therefore  one should consider the simplest configuration:
electrons gyrate in the plane perpendicular to the background magnetic field, and the wave
 polarization vector, as well as the propagation direction, lie in the same plane.
 Let the wave propagate in the $x$ direction and be polarized in the
$y$ direction whereas the background magnetic field be directed in $z$ direction.
 Then the wave is described by the vector potential
 \be
 \mathbf{A}=\frac{cE}{\omega}\cos\eta\mathbf{\widetilde y};\\
\eta=\omega(t-x/c);\label{eta}
\ee
and the electron equations of motion are written as
\be
mc\frac{du_x}{dt}&=&\frac{ev_y}c\left(\frac{\partial A}{\partial x}+B_0\right);\label{ux}\\
mc\frac{du_y}{dt}&=&-\frac ec\left(\frac{\partial A}{\partial t}+v_x\frac{\partial A}{\partial
x}+v_xB_0\right); \label{uy}\\
mc^2\frac{d\gamma}{dt}&=&-\frac{ev_y}c\frac{\partial A}{\partial t}.\label{gamma}
\ee
where $\gamma=(1-v^2/c^2)^{-1/2}$ is the electron Lorentz factor, $\mathbf{u}=(\mathbf{v}/c)\gamma$
the 4-velocity, $B_0$ the background magnetic field. The electron is assumed to move in the $x-y$
plane.

It is well known that if there is no the background field, $B_0=0$, the above system of equations
has two integrals of motion (e.g., \citealp{Gunn_Ostriker71,Landau_Lifshitz75}). Invariancy with
respect to a shift in the $y$ direction implies conservation of the $y$ component of the
generalized momentum, which means that the quantity
\begin{equation}
w=u_y+a\cos\eta \label{u_y}
\end{equation}
remains constant. One sees that if $a> 1$, the electron oscillations
  become relativistic.  Invariancy with respect to a
transformation $x\to x+s,\, t\to t+s$ implies conservation of the quantity
\eqb
g=\gamma-u_x. \label{G}
\eqe
Making use of the identity $u_x^2+u_y^2+1=\gamma^{2}$, one expresses the velocity components and
the electron Lorentz factor  via the integrals of motion as
\be
v_x=\frac{1+(w-a\cos\eta)^2-g^2}{1+(w-a\cos\eta)^2+g^2}c;\label{vx}\\
v_y=\frac{2g(w-a\cos\eta)}{1+(w-a\cos\eta)^2+g^2}c; \label{vy}\\
 \gamma=\frac{1+(w-a\cos\eta)^2+g^2}{2g}.\label{energy}
\ee
The relation between the time, $t$, and the phase, $\eta$, is found by differentiating eq.
(\ref{eta}) with respect to $t$ and using eq. (\ref{G}):
\eqb
\frac{d\eta}{dt}=\frac{\omega g}{\gamma}. \label{eta-t}
\eqe
The electron "sees" the full period of the wave for the time
\eqb
T=\frac 1{\omega g}\int_0^{2\pi}\gamma d\eta=\frac{\pi}{\omega g^2}\left(1+w^2+\frac
12a^2+g^2\right).
\eqe
The  components of the velocity and the Lorentz factor averaged over the wave period are found as
\be
\overline{v_x}&=&\frac 1T\int_0^Tv_xdt=\frac c{\omega Tg}\int_0^{2\pi}u_x d\eta=
\frac{1+w^2+\frac 12a^2-g^2}{1+w^2+\frac 12a^2+g^2}c;\label{vx_average}\\
\overline{v_y}&=&\frac{2wg}{1+w^2+\frac 12a^2+g^2}c; \label{vy_average}\\
\overline{\gamma}&=&\frac{(1+g^2+w^2+\frac 12a^2)^2+2a^2w^2+\frac 38a^4}{2g\left(1+w^2+g^2+\frac
12a^2\right)}.\label{gamma_average}
\ee
The velocity of the electron guiding centre is written as
\eqb
\left(\overline{v}\right)^2=\left(\overline{v_x}\right)^2+\left(\overline{v_y}\right)^2=
c^2-\frac{4g^2\left(1+\frac 12a^2\right)}{\left(1+w^2+\frac 12a^2+g^2\right)^2}c^2.
\eqe

In the presence of the background magnetic field, one can find the electron motion if the wave
frequency is large as compared with the Larmor frequency,
 \be
 \omega\gg\omega_B\equiv\frac{eB_0}{mc};
 \ee
  then the electron
motion could be described as rapid oscillations superimposed on a slow Larmor rotation of
the guiding centre. In this case, one can conveniently use the "integrals of motion", $g$
and $w$, as new unknowns. Differentiating eqs. (\ref{u_y}) and (\ref{G}) in time and
making use of eqs. (\ref{ux}) and (\ref{uy}) yields
\be
\frac{dg}{dt}=-\omega_Bv_y; \\
\frac{dw}{dt}=-\omega_Bv_x.
\ee
 Now making use of eqs. (\ref{vx}), (\ref{vy}), (\ref{energy}) and (\ref{eta-t}), one gets the
closed system of equations
\be
g\frac{dg}{d\eta}&=&-\frac{\omega_B}{\omega}(w-a\cos\eta); \label{eq1}\\
g^2\frac{dw}{d\eta}&=&-\frac{\omega_B}{2\omega}\left[1+(w-a\cos\eta)^2-g^2\right].\label{eq2}
\ee
One sees that one can use $g^2$ instead of $g$ as an unknown function. These equations could be
solved by separating slow and rapid motions.

\section{Motion averaged over the rapid oscillations}

Let us present the unknown functions in the form
\eqb
g^2=G^2+\psi;\quad w=U+\xi; \label{expansion}
\eqe
where 
$G$ and $U$ are 
a slowly varying quantities defined as $G^2=(2\pi)^{-1}\int_0^{2\pi}g^2d\eta$ and
$U=(2\pi)^{-1}\int_0^{2\pi}w^2d\eta$ whereas $\psi$ and $\xi$ are  small rapidly oscillating
corrections.
Substituting this expansion into 
eqs. (\ref{eq1}) and (\ref{eq2}), linearizing in small $\psi$ and $\xi$ and averaging in
$\eta$ yields equations describing motion of the guiding centre:
\be
\frac{dG^2}{d\eta}&=&-\frac{2\omega_B U}{\omega}; \label{average1}\\
\frac{dU}{d\eta}&=&-\frac{\omega_B}{2\omega G^2}\left[1+U^2+\frac 12a^2-G^2\right].\label{average2}
\ee
Dividing the second equation by the first one, one gets a linear equation with respect to $U^2$:
\eqb
2G\frac{dU^2}{dG}=U^2+1+\frac 12a^2-G^2,
\eqe
which is solved giving the first integral of 
the system (\ref{average1}) and (\ref{average2}):
\eqb
U^2+(G-\Gamma)^2=\Gamma^2-1-\frac 12a^2, \label{circle}
\eqe where $\Gamma$ is a constant.

According to the above solution, the electron moves along a circle in the $U-G$ plane.
The motion of the guiding centre in the coordinate space is described by eqs.
(\ref{vx_average}) and (\ref{vy_average}). Substituting $w$ and $g$ by  $U$ and $G$,
correspondingly, and making use of eq. (\ref{circle}) yields
\eqb
\frac{\overline{v_x}}c=1-\frac G{\Gamma};\quad \frac{\overline{v_y}}c=\frac U{\Gamma}; \quad
\frac{\overline{v}}c=\frac{\sqrt{\Gamma^2-1-\frac 12a^2}}{\Gamma}.
\eqe
 One sees that the guiding centre of the electron gyrates around the magnetic field with a constant velocity;
 the Lorentz factor of the averaged motion is found as
\eqb
\gamma_{\rm gc}=\frac 1{\sqrt{1-\overline{v}^2/c^2}}=\frac{\Gamma}{\sqrt{1+\frac 12a^2}}.
\label{guid_center_Lorentz}\eqe
  The  averaged velocity is relativistic if
\eqb
\Gamma^2\gg 1+\frac 12a^2; \label{condition}
\eqe
below this condition is assumed to be fulfilled. 

The variable $U$, which is the averaged vertical component of the electron 4-velocity, varies from
$U=-\sqrt{\Gamma^2-1-\frac 12a^2}\approx -\Gamma$ to $U=\sqrt{\Gamma^2-1-\frac 12a^2}\approx
\Gamma$ and vanishes twice during the rotation period, at the upper and the lower points of the
electron orbit. The electron moves in the direction of the wave in the upper part of the orbit and
towards the wave in the lower point. At these points, $G$ reaches minimum and maximum,
correspondingly:
\eqb
G_{\rm min,\,max}=\Gamma\pm\sqrt{\Gamma^2-1-\frac 12a^2}\approx\left\{\begin{array}{l}
 \frac{1+\frac 12a^2}{2\Gamma};
\\ 2\Gamma.\end{array}\right.\label{Gmin}
\eqe
 The averaged Lorentz factor, eq. (\ref{gamma_average}), varies along the Larmor orbit as
 \eqb
\frac{\overline{\gamma}}{\Gamma}=1+\left(\frac a{\Gamma}\right)^2\frac{U^2+\frac 3{16}a^2}{2G^2}.
 \label{av-Lorentz}\eqe
One sees that in the case of weak waves, $a\ll 1$, the constant $\Gamma$ is just the Lorentz factor
of the electron. Inspection of eqs. (\ref{av-Lorentz}) and (\ref{circle}) shows that for strong
waves at the condition (\ref{condition}), $\overline{\gamma}$ remains close to $\Gamma$ in the most
of the Larmor orbit and only in the upper part increases reaching
\eqb
\overline{\gamma}_{\rm max}=\left(1+\frac{3a^4}{2(2+a^2)^2}\right)\Gamma
\eqe
in the upper point. Beyond the upper point, $\overline{\gamma}$ decreases and goes to $\Gamma$
again therefore when considering only motion averaged over the rapid oscillations, one could not
find the net energy gain due to the absorption of the wave. One has to find the corrections $\psi$
and $\xi$, which will be done in the next section.

In order to find the dependence of the variables on time, one can use eq. (\ref{eta-t}).
Substituting $\gamma$ from eq. (\ref{energy}) and integrating, one gets the relation between the
phase and the time; for $\eta\gg 1$ it looks like
 \eqb
t=\int\frac{1+(w-a\cos\eta)^2+g^2}{2g^2}d\eta.
 \eqe
Neglecting oscillating parts of $w$ and $g$,  one can substitute them by $U$ and $G$,
correspondingly. Substituting the other oscillating terms by their averaged values, one gets
 \eqb
t=\int\frac{1+w^2+\frac 12a^2+g^2}{2g^2}d\eta=-\frac 1{\omega_B}\int\frac{1+U^2+\frac
12a^2+G^2}{UG}dG.
 \eqe
 Here in the last equality, the integration variable has been substituted by
$G$ with the aid of eq. (\ref{average1}). The integral is performed after expressing $U$ via $G$
with the aid of eq. (\ref{circle}); then one finds
 \be
G&=&\Gamma-\left(\Gamma^2-1-\frac 12a^2\right)^{1/2}\cos\Omega(t-t_0);\\
\Omega&=&\frac{\omega_B}{\Gamma}.
 \label{Larmor}\ee
 One sees that in the presence of a strong wave, the guiding centre of the electron experiences Larmor
 rotation around the background magnetic field. Taking into account that the constant $\Gamma$ is equal
 to the average
 Lorentz factor in the most of the orbit, the Larmor period is not affected by the wave.

\section{Oscillations with respect to the averaged motion}
 In order to find
oscillations with respect to the average motion of the guiding centre, one linearizes
eqs. (\ref{eq1}) and (\ref{eq2})  in small $\psi$ and $\xi$ and eliminates the zeroth
order terms by extracting eqs. (\ref{average1}) and (\ref{average2}); this yields a set
of equations
\be\frac{d\psi}{d\eta}=\frac{2\omega_B}{\omega}(a\cos\eta-\xi); \\
G^2\frac{d\xi}{d\eta}+\psi\frac{dU}{d\eta}=-\frac{\omega_B}{\omega}\left\{U\xi
-aU\cos\eta+\frac{a^2}4\cos2\eta-\frac 12\psi\right\}.
 \ee
Eliminating $\xi$ and making use of eq. (\ref{average2}), one gets a single equation for $\psi$
\be
\frac{d^2\psi}{d\eta^2}+\frac{\omega_B}{\omega}\frac{U}{G^2}\frac{d\psi}{d\eta}+
\frac{\omega_B^2}{\omega^2G^4}\left(1+\frac 12a^2+U^2\right)\psi\nonumber\\
=-\frac{2\omega_Ba}{\omega}\sin\eta+\frac{\omega^2_Ba^2}{2\omega^2G^2}\cos2\eta.
\label{linear-psi}\ee
 In this equation, $U$ and $G$  are related by eq. (\ref{circle}); the dependence of these functions on $\eta$
 may be found by substituting eq. (\ref{circle}) into eq. (\ref{average1}).

The relativistic electron exchanges energy with the wave in the upper part of the orbit, where it
moves in the direction of the wave thus remaining for a long time in phase with the wave. It
follows from eqs. (\ref{condition}) and (\ref{Gmin}) that in the upper part of the orbit, $G\ll
\Gamma$; then eq. (\ref{circle}) is reduced to
\eqb
G=\frac{1+\frac 12a^2+U^2}{2\Gamma}. \label{G-upper}
\eqe
Substituting this relation into eq. (\ref{average1}) and integrating, one gets
\eqb
\eta=\phi-\frac{\omega}{2\omega_B\Gamma^2}\left[\left(1+\frac
12a^2\right)U+\frac{U^3}3\right],\label{eta-upper}
\eqe
 where $\phi$ is the phase of the wave when the electron passes the upper point of the orbit. Eqs.
(\ref{G-upper}) and (\ref{eta-upper}) describe, in parametric form, motion of the
electron guiding centre in the upper part of the orbit. Recall that $U$ is the averaged
over rapid oscillations vertical component of the electron 4-velocity; it passes zero at
the upper point of the orbit.

Let us now solve eq. (\ref{linear-psi}) in the upper part of the orbit.
Instead of substituting directly eqs. (\ref{G-upper}) and
(\ref{eta-upper}) into the equation, one can conveniently introduce a new independent variable
 \eqb
z=-\frac{U}{\sqrt{1+\frac 12a^2}}.
 \label{z}\eqe
Then
 \be
 \eta=\phi+S\left[z+\frac{z^3}3\right];\label{eta-z}\\
S=\frac{\omega}{2\omega_B\Gamma^2}\left(1+\frac 12a^2\right)^{3/2};\label{S}\\
 G=\left(1+\frac 12a^2\right)\frac{1+z^2}{2\Gamma}
 \label{G-z}\ee
Now eq. (\ref{linear-psi}) takes the form
 \be (1+z^2)\frac{d^2\psi}{dz^2}-4z\frac{d\psi}{dz}+4\psi\nonumber\\=
- \frac{aS\left(1+\frac
12a^2\right)^{3/2}}{\Gamma^2}\left\{\left(1+z^2\right)^3
\sin\left[\phi+S\left(z+\frac{z^3}3\right)\right]\right.
\label{psi-eq}\\\left.-\frac{a(1+z^2)}{2S\left(1+\frac
12a^2\right)}\cos2
\left[\phi+S\left(z+\frac 13z^3\right)\right]\right\}.\nonumber \ee
 One can check easily that the corresponding homogeneous equation is satisfied by $\psi=z$ and
$\psi=1-2z^2-\frac 13z^4$. Then  variation of constants yields the solution of eq. (\ref{G-z}) in
the form
 \be
\psi = \frac{aS\left(1+\frac 12a^2\right)^{3/2}}{\Gamma^2}
\int_{-\infty}^z(z'-z)\left[1+2zz'+\frac{zz'}3(z^2+zz'+z'^2)\right] \nonumber\\\
\times\left\{\sin\left[\phi+S\left(z'+\frac{z'^3}3\right)\right]
\label{solution1}\right.\\\left.-\frac{a}{2S\left(1+\frac
12a^2\right)^{1/2}(1+z'^2)^2}\cos2
\left[\phi+S\left(z'+\frac 13z'^3\right)\right]\right\}dz'. \nonumber\ee

\section{The energy exchange between the electron and the wave}

Variation of the particle energy could be found by differentiating eq. (\ref{energy}) for the
particle Lorentz factor and making use of eqs. (\ref{eq1}) and (\ref{eq2}):
\eqb
\frac{d\gamma}{d\eta}=-\frac{\omega a}{\omega_B}\frac{dg}{d\eta}\sin\eta. \label{dgamma}
\eqe
 The energy gain  after passing the upper part of the orbit is found by
integrating eq. (\ref{dgamma}).  In all practical cases, the absorption by an ensemble of
homogeneously disributed electrons is of interest; therefore the result should be averaged in
phases, $\langle\dots\rangle=(2\pi)^{-1}\int_0^{2\pi}\dots d\phi$. Performing integration by parts,
one gets
\eqb
\langle\Delta\gamma\rangle=-\frac{\omega a}{\omega_B}\left\langle\int_{-\infty}^{\infty}
\frac{dg}{d\eta}\sin\eta d\eta\right\rangle =\frac{\omega a}{\omega_B}
\left\langle\int_{-\infty}^{\infty}g\cos\eta d\eta\right\rangle.
\eqe
In order to get a non-zero result after averaging in phases, one has to take into account
oscillations of the electron with respect to the guiding centre.

It follows from the expansion (\ref{expansion}) that
\eqb
g=\sqrt{G^2+\psi}=G+\frac{\psi}{2G}.
\eqe
Then the particle energy gain is written as
 \be
  \langle\Delta\gamma\rangle=\frac{\omega
a}{2\omega_B}\left\langle\int_{-\infty}^{\infty} \psi\cos\eta
\frac{d\eta}G\right\rangle\nonumber\\= \frac{a\omega^2}{2\omega_B^2\Gamma}\left(1+\frac
12a^2\right)^{1/2} \int_{-\infty}^{\infty}\left\langle \psi\cos\left[\phi+S\left(z+\frac
13z^3\right)\right]\right\rangle dz
 \label{delta_gamma}\ee
where in the last equality, eqs. (\ref{eta-z}), (\ref{S}) and (\ref{G-z}) were used.

When substituting the solution (\ref{solution1}) into eq. (\ref{delta_gamma}), the term with $\cos
2\left[\phi+S\left(z'+\frac 13z'^3\right)\right]$ vanishes after the averaging in phases. The term
with $\sin\left[\phi+S\left(z'+\frac 13z'^3\right)\right]$ is transformed as
\eqb
\langle\sin\eta'\cos\eta\rangle=\frac 12\sin(\eta'-\eta)=\frac 12\sin\left[S\left(z'-z+\frac
13z'^3-\frac 13z^3\right)\right].
\eqe
Then one finds
\be
\langle\Delta\gamma\rangle=\frac{a^2\omega^3}{4\omega_B^3\Gamma^5}\left(1+\frac 12a^2\right)^{7/2}
\int_{-\infty}^{\infty}dz\int_{-\infty}^{z}(z'-z)
\left[1+2zz'\right.\nonumber\\\left.+\frac{zz'}3(z^2+zz'+z'^2)\right]
\sin\left[S\left(z'-z+\frac{z'^3-z^3}3\right)\right]dz'.
\ee
The integrand is symmetric with respect to exchange $z$ and $z'$, therefore one can extend the
integration domain to the whole $z-z'$ plane, $\int_{-\infty}^{\infty}dz\int_{-\infty}^{z}\dots
dz'=\frac 12 \int_{-\infty}^{\infty}dz\int_{-\infty}^{\infty}\dots dz'$. Then the double integral
is split into two 1D integrals. Taking into account parity of functions, one gets
\be
\langle\Delta\gamma\rangle=\frac{a^2\omega^3}{2\omega_B^3\Gamma^5}\left(1+\frac 12a^2\right)^{7/2}
\int_{0}^{\infty}z\sin\left[S\left(z+\frac{z^3}3\right)\right]dz \nonumber\\
\times\int_{0}^{\infty}\left(1-2z^2-\frac
13z^4\right)\cos\left[S\left(z+\frac{z^3}3\right)\right]dz\label{delta_g1}\\
=\frac{8\pi^2 a^2 S^{2/3}\Gamma}{3\left(1+\frac 12a^2\right)}
\rm{Ai'}\,(S^{2/3})[\rm{Ai'}\,(S^{2/3})-4S^{4/3}\rm{Ai}\,(S^{2/3})]; \label{delta_g}
 \ee
 where
\eqb
\rm{Ai}\,(t)=\frac 1{\pi}\int_0^{\infty}\cos\left(tx+\frac 13x^3\right)dx
\eqe
is the Airy function. One sees that the range of $z$ satisfying the condition
$S\left(z+\frac{z^3}3\right)\sim1$, which corresponds, according to eq. (\ref{eta-z}), to
$\eta\sim1$, contributes to the integrals. This means that the electron gains energy at $\eta\sim
1$, i.e. when it moves in phase with the wave at the upper part of the orbit.

At a small or a large $S$, one finds simple relations
\eqb
\langle\Delta\gamma\rangle=\frac{8\pi a^2\Gamma}{3\left(1+\frac
12a^2\right)}\left\{\begin{array}{ll} \frac{\pi }{3^{2/3}[\Gamma(1/3)]^2}S^{2/3};\quad S\ll 1;
\\ S^{2}e^{-\frac 43S};\quad S\gg 1;\end{array}\right.
\eqe
where $\Gamma(x)$ is the gamma-function. One sees that the particle energy gain in one Larmor
period is maximal at $S\sim 1$. In a weak wave, $a\ll 1$, the energy gain is always small,
$\langle\Delta\gamma\rangle\ll 1$. In a strong wave, $a\geq 1$, it becomes significant,
$\langle\Delta\gamma\rangle\sim\Gamma$, at $S\sim 1$.

The absorption cross-section, $\sigma$, is defined such that the energy absorbed by an electron per
unit time is $\sigma$ times the Poynting flux in the wave. The electron absorbs on the average the
energy $mc^2\langle\Delta\gamma\rangle$ per rotation period, $T_B=2\pi\Gamma/\omega_B$, therefore
one can write
\eqb
mc^2\langle\Delta\gamma\rangle\frac{\omega_B}{2\pi\Gamma}=\sigma\frac{E_0^2}{8\pi}c.
\eqe
Substituting eq. (\ref{delta_g}) yields finally
\be
\sigma=\frac{2^{13/3}\pi r_ec\omega_B^{1/3}}{3\omega^{4/3}\Gamma^{4/3}}
\rm{Ai'}\,\left[(\omega/\omega_0)^{2/3}\right]\nonumber\\\times\left\{\rm{Ai'}\,\left[(\omega/\omega_0)^{2/3}\right]-
4(\omega/\omega_0)^{4/3}\rm{Ai}\left[(\omega/\omega_0)^{2/3}\right]\right\} \label{sigma}\\=
\frac{2^{7/3}\pi^2 r_ec\omega_B^{1/3}}{3\Gamma^{4/3}}\left\{\begin{array}{ll} \frac{4\pi
}{3^{2/3}[\Gamma(1/3)]^2}\omega^{-4/3};\quad \omega\ll \omega_0;
\\ \frac{\omega^{2/3}}{\omega_0^2}e^{-\frac{4\omega}{3\omega_0}};\quad \omega\gg \omega_0;\end{array}\right.,
 \ee
 where
 \eqb
\omega_0=\frac{2\omega_B\Gamma^2}{\left(1+\frac 12a^2\right)^{3/2}}\, ,\label{omega0}
\eqe
$r_e=e^2/mc^2$ is the classical electron radius. At a small $a$, this expression reduces to the
classical expression for the synchrotron absorption (see Appendix).

\section{Validity of the perturbative solution}

The above results are based on the perturbative solution, which is valid if $\psi\ll G^2$. In the
upper part of the electron trajectory, where the electron exchanges energy with the wave, $G$ is
small  
 therefore this condition could be
violated. In order to check validity of the obtained solution, let us estimate $\psi$ directly from
eq. (\ref{linear-psi}), which is simpler than finding estimates from the exact solution
(\ref{solution1}). Substituting eq. (\ref{G-z}) and (\ref{z}), one can write this equation as
\be
\frac{d^2\psi}{d\eta^2}-\frac{2z}{S(1+z^2)^2}\frac{d\psi}{d\eta}+ \frac{4}{S^2(1+z^2)^3}\psi\\
=-\frac{2\omega_Ba}{\omega}\left(\sin\eta+\frac{a}{2S\left(1+\frac 12a^2\right)^{1/2}(1+z^2)^2}\cos2\eta\right).
\nonumber \ee
 Taking into account that $\frac{d\psi}{d\eta}\sim\psi$, one could estimate $\psi$ just balancing
 terms in the equation.

 At $S\gg 1$, the lhs of the equation is dominated by the first term and the rhs is dominated by the first term.
 Therefore $\psi\sim \omega_Ba/\omega$, which implies
 \eqb
\frac{\psi}{G^2}\sim \frac a{\left(1+\frac 12a^2\right)^{1/2}S(1+z^2)^2}\ll 1.
 \label{large_S}\eqe
One sees that at large $S$, i.e. at large frequencies, $\psi$ remains small as compared with $G^2$
at any $a$. It is no surprise that the approximate solution is valid in this case, because it
follows from eq. (\ref{eta-z}) that at $S\gg 1$, the phase of the wave, $\eta$, rapidly varies when
$z$, and therefore $U$ and $G$, vary slowly, which was an initial assumption of our perturbation
method.

Now let us consider the case $S\ll 1$. In this case, we have to consider a few ranges of $z$
separately. If $z\geq S^{-1/3}$, both the lhs side and the rhs of the equation are dominated by
their first terms therefore the estimate (\ref{large_S}) remains valid in this case too. In the
small range $S^{-1/4}<z<S^{-1/3}$, the lhs is dominated by the last term whereas the rhs is still
dominated by the first term therefore one finds
\eqb
\psi\sim\frac{a\omega_B}{\omega}S^2z^6,
\eqe
which yields
\eqb
\frac{\psi}{G^2}\sim \frac a{\left(1+\frac 12a^2\right)^{1/2}}Sz^2\ll 1.
\eqe
In the case $z<S^{-1/4}$, both the lhs and the rhs of the equation are dominated by their last
terms; then
\eqb
\psi\sim\frac{a^2\omega_B}{\omega\left(1+\frac 12a^2\right)^{1/2}}S(1+z^2),
\eqe
and
\eqb
\frac{\psi}{G^2}\sim \frac{a^2}{\left(1+\frac 12a^2\right)^{1/2}(1+z^2)}.
\eqe
One sees that in the case of weak wave, $a\ll 1$, the condition $\psi\ll G^2$ is fulfilled at any
$z$ therefore our approximation is valid everywhere. In the case of strong waves, $a\geq 1$, it is
valid in the most of the Larmor orbit with the exception of a region $z\sim 1$, where $\psi\sim
G^2$. Note that the energy exchange between the wave and the electron occurs at $\eta\sim 1$. For
$S\ll 1$, this corresponds to $z\sim S^{-1/3}\gg 1$; it is this range of $z$ that contributes to
integrals in eq. (\ref{delta_g1}). Taking into account that our approximation is valid at $z\gg 1$
and
 is marginally fulfilled $z\sim 1$, one concludes that the expressions for the particle energy gain
 and for the absorption coefficients, eqs. (\ref{delta_g1})-(\ref{sigma}), are valid at $S\ll 1$.

Now let us consider the case $S\sim 1$. Then $\psi\sim a\omega_B/\omega$ and
\eqb
\frac{\psi}{G^2}\sim \frac a{\left(1+\frac 12a^2\right)^{1/2}(1+z^2)^2}.
\eqe
One sees that  for weak waves, $a\ll 1$, our approximation is valid at any $z$ whereas for strong
waves, it becomes marginally correct at $z\sim 1$. At $S\sim 1$, the electrons gain energy at
$z\sim 1$ therefore one finally concludes that the expression (\ref{sigma}) for the synchrotron
absorption cross-section is always correct for weak waves whereas for strong waves, it is correct
in the high and low frequency limits, $S\gg 1$ and $S\ll 1$, and could be used as an estimate for
$S\sim 1$.

\section{Discussion}

It is well known that the synchrotron emission and absorption occur in a wide frequency range at
high harmonics of the rotation period. According  to the standard theory (e.g.,
\citealp{Landau_Lifshitz75,Melrose80}), the characteristic frequency is
$\omega_0\sim\omega_B\gamma^2$, which corresponds to $\sim\gamma^3$
harmonics. 
It was shown in this paper that for strong waves, the synchrotron absorption occurs at high
harmonics too but the characteristic  frequency significantly decreases, see eq. (\ref{omega0}),
therefore the frequency range for an efficient absorption also decreases significantly. Let us
discuss the physical origin of the phenomenon.

The rotation frequency in the presence of a high-frequency electromagnetic wave is given by eq.
(\ref{Larmor}), where $\Gamma$ is the constant of motion, which is equal to the particle Lorentz
factor if the wave is weak; in a strong wave, it is equal to the average particle Lorentz factor in
the most of the orbit. Therefore the rotation period is not affected by the wave. On the other
hand, the rotational motion averaged over the rapid oscillations occurs with the Lorentz factor
(\ref{guid_center_Lorentz}), which is significantly smaller than $\Gamma$ for a large strength
parameter $a$.

The phase of the wave seen by the electron guiding centre, eq. (\ref{eta}), is
 \eqb
 \eta=\omega\left(t-\frac{\overline{v}}{\Omega}\sin\Omega t\right).
 \eqe
The electron absorbs radiation at a small fraction of the trajectory, where it moves together with
the wave thus remaining relatively long time in phase with the wave. Then the frequency "seen" by
the electron, $\frac{d\eta}{dt}$,  is minimum; in our case, this occurs near the upper point of the
orbit, $t=0$. Expanding around this point yields
 \eqb
\eta=\omega\left(\frac t{2\gamma_{\rm gc}^2}+\frac 16\Omega^2t^3\right)=
\frac{\omega}{2\Omega\gamma_{\rm gc}^3}\left[\gamma_{\rm gc}\Omega t+\frac 13(\gamma_{\rm gc}\Omega
t)^3\right].
 \eqe
One now sees that the characteristic synchrotron frequency is $\omega_0\sim\Omega\gamma_{\rm
gc}^3$, which reproduces formula (\ref{omega0}). At $\omega\leq\omega_0$, the electron exchanges
energy with the wave during the time interval $\Omega t\sim (\omega/\Omega)^{1/3}$ corresponding to
$\Delta\eta\sim 1$; beyond this interval, the oscillation frequency rapidly grows so that the
energy exchange does not occur on the average. At $\omega\gg\omega_0$, the electron experiences a
few oscillations while $\gamma_{\rm gc}\Omega t<1$, when the wave frequency, $\frac{d\eta}{dt}$,
remains constant; then the average energy exchange is small.

Other absorption mechanisms, as well as the application of the obtained results to the termination
shocks in PWNe, will be discussed in the next papers of the series.

 This research was supported by the grant I-1362-303.7/2016 from the German-Israeli Foundation
for Scientific Research and Development.

\section*{Appendix.Syncrotron absorption of weak waves}
As a consistency check, let us find the absorption coefficient of weak waves for the configuration
used in this paper. This could be conveniently done by making use of the Einstein coefficient
method. The evolution of the photon occupation number, $n_{\bf k}$, is governed by the kinetic
equation, which is written with account of the detailed balance principle in the form
\eqb
\frac{\partial n_{\mathbf k}}{\partial t}=\int W(\mathbf{p,k})\left\{f_{\mathbf{p}}(1+n_{\mathbf
k})-n_{\mathbf k}f_{\mathbf p-\hbar \mathbf{k}}\right\}\frac{d^3p}{(2\pi\hbar)^3},
\eqe
where $f_{\mathbf p}$ is the electron distribution function, $W({\mathbf p,k})$ the probability for
spontaneous emission of a photon with the wave vector $\mathbf k$ by an electron with the momentum
$\mathbf p$. We are interested in synchrotron emission/absorption of highly relativistic electrons
rotating perpendicularly to the magnetic field therefore the element of the phase volume may be
conveniently written in the cylindrical coordinates as $d^3p=c^{-2}\varepsilon d\varepsilon
d\varphi dp_z$, where $\varepsilon=cp$ is the electron energy,  whereas the electron distribution
function may be presented as
 \eqb
 f_{\mathbf p}=(2\pi c)^2\hbar^3\frac{N(\varepsilon)}{\varepsilon}\delta(p_z),
 \eqe
 where $N(\varepsilon)$ is the number density of electrons per unit energy range.

We consider radiation in the plane $k_z=0$; in this case, the emission probability depends on the
electron energy, $\varepsilon$, the photon frequency, $\omega$, and the angle $\theta$ between
$\mathbf p$ and $\mathbf k$. A highly relativistic electron radiates in the direction of motion
therefore one can write $W({\mathbf p,\mathbf k})=2\pi
Y(\varepsilon,\omega)\delta(\varphi-\varphi')$, where the angle $\varphi'$ shows the direction of
the photon in the $x-y$ plane, $\tan\varphi'=k_y/k_x$. Then the kinetic equation is written as
\eqb
\frac{\partial n_{\mathbf k}}{\partial t}=\int_0^{\infty}
Y({\varepsilon,\omega})\left\{N(\varepsilon)(1+n_{\mathbf k})- n_{\mathbf
k}\frac{\varepsilon}{\varepsilon-\hbar\omega}N(\varepsilon-\hbar\omega)\right\} d\varepsilon
\eqe
Instead of the photon occupation number, $n_{\mathbf k}$, one can conveniently use the radiation
intensity,
 \eqb
 I=\frac{\hbar\omega^3}{(2\pi c)^{3}} n_{\mathbf k}.
 \eqe
 Substituting $n_{\mathbf k}$ by $I$ and expanding
 in small $\hbar\omega\ll\varepsilon$, one reduces the kinetic equation to the standard
form of the radiation transfer equation
\eqb
\frac{\partial I}{\partial t}=j-\kappa I, \label{rad_transfer}
 \eqe
 where
 \eqb
j=\frac{\hbar\omega^3}{(2\pi c)^3}\int_0^{\infty} Y({\varepsilon,\omega})N(\varepsilon)d\varepsilon
 \label{j}\eqe
 is the emissivity and
 \be
\kappa=-\hbar\omega\int_0^{\infty} Y({\varepsilon,\omega})\varepsilon\frac
d{d\varepsilon}\left(\frac{N(\varepsilon)}{\varepsilon}\right)d\varepsilon\nonumber\\
=\hbar\omega\int_0^{\infty}\frac{N(\varepsilon)}{\varepsilon}\frac{d}{d\varepsilon}
\left(\varepsilon Y({\varepsilon,\omega})\right) d\varepsilon
\ee
 the absorption coefficient. Now the absorption cross-section may be presented as
\eqb
\sigma=\frac{\hbar\omega}{\varepsilon}\frac{d}{d\varepsilon}\varepsilon Y({\varepsilon,\omega}),
 \label{absorp_linear}\eqe
 i.e. the absorption is related to the spontaneous emission power.

The radiation of an electron gyrating perpendicularly to the magnetic field is calculated, e.g., in
\citet{Landau_Lifshitz75}. The electron radiates in harmonics of the rotation frequency,
$\omega=n\omega_B/\gamma$. The emission power in the rotation plane at the $n$-th harmonic is found
as
\eqb
dI_n=\frac{n^2e^2\omega_B^2v^2}{2\pi c\gamma^2c^2}J'^2_n\left(\frac{nv}c\right)d\Omega,
\eqe
where $J'_n(x)$ is the derivative of the Bessel function of $n$-th order. For high harmonics, $n\gg
1$, the emission power in a frequency interval $d\omega=(\omega_B/\gamma)dn$ is presented as
 \eqb
dP=dI_n\frac{\gamma}{\omega_B}d\omega.
 \eqe
On the other hand, it follows form eq. (\ref{j}) that the single electron  emission power is
presented as
 \eqb
dP=\frac{\hbar\omega^3}{(2\pi c)^3}Yd\omega d\Omega.
 \eqe
Comparing these two expressions, one finds
\eqb
Y=\frac{4\pi^2e^2\gamma v^2}{\hbar\omega\omega_B}J'^2_n\left(\frac{nv}c\right).
\eqe
In the case of interest, $v\approx c$, $n\gg 1$, one can use the asymptotic relation
 \eqb
J_n(n\xi)=\left(\frac 2n\right)^{1/3}{\rm Ai}\left[2^{1/3}n^{2/3}(1-\xi)\right];\quad n\gg1;\quad
\xi\approx 1;
 \eqe
 to yield
 \eqb
Y=\frac{2^{10/3}\pi^2e^2c^2\omega_B^{1/3}}{\hbar\omega^{7/3}\gamma^{1/3}} {\rm
Ai'}^2\left(R^{2/3}\right);\quad R=\frac{\omega}{2\omega_B\gamma^2}.
 \eqe
Substituting this expression into eq. (\ref{absorp_linear}), one gets the synchrotron absorption
cross-section in the form
 \eqb
 \sigma=\frac{2^{13/3}\pi^2e^2c^2r_e\omega_B^{1/3}}{3\omega^{4/3}\gamma^{4/3}}
{\rm Ai'}\left(R^{2/3}\right)\left[{\rm Ai'}\left(R^{2/3}\right)-4R^{4/3}{\rm
Ai}\left(R^{2/3}\right)\right].
 \eqe
One sees that this expression coincides with eq. (\ref{sigma}) at $a\ll 1$, when $R=S$.

\bibliographystyle{mn2e}

\end{document}